# Optimization of Tensor-product Operations in Nekbone on GPUs

Martin Karp, Niclas Jansson, Artur Podobas, Philipp Schlatter, and Stefano Markidis
*KTH Royal Institute of Technology*
Stockholm, Sweden
martin.karp@outlook.com, njansson@kth.se, podobas@kth.se, markidis@kth.se

*Abstract*—In the CFD solver Nek5000, the computation is dominated by the evaluation of small tensor operations. Nekbone is a proxy app for Nek5000 and has previously been ported to GPUs with a mixed OpenACC and CUDA approach. In this work, we continue this effort and optimize the main tensor-product operation in Nekbone further. Our optimization is done in CUDA and uses a different, 2D, thread structure to make the computations layer by layer. This enables us to use loop unrolling as well as utilize registers and shared memory efficiently. Our implementation is then compared on both the Pascal and Volta GPU architectures to previous GPU versions of Nekbone as well as a measured roofline. The results show that our implementation outperforms previous GPU Nekbone implementations by 6-10%. Compared to the measured roofline, we obtain 77 - 92% of the peak performance for both Nvidia P100 and V100 GPUs for inputs with 1024 - 4096 elements and polynomial degree 9.

## I. INTRODUCTION

Nek5000 [1] is a high order solver for Computational Fluid Dynamics (CFD) based on the Spectral Element Method (SEM) that solves the Navier-Stokes equation for incompressible flow and has been in use for over 30 years (Gordon Bell Prize 1999 [2]). In its current form, the solver uses conventional processors and legacy Fortran 77 combined with C and MPI for parallelization. As we are moving to exa-scale computations and more heterogeneous hardware (several of the top 10 supercomputers in the world use both GPUs and CPUs [3]) a major effort to modernize Nek5000 and port it to GPUs and other accelerators is necessary. Nekbone [4] is a proxy app for Nek5000 that illustrates important computational and scaling aspects of the entire solver. If we can make improvements in Nekbone, these can be transferred to Nek5000. In particular, the evaluation of the Poisson operator through a tensor-product operation is the most time consuming part of both Nekbone and Nek5000 [5]. Therefore, optimizing the tensor-product is one of the most important steps towards a high performing version of Nek5000 on GPUs. In this work, we research how the tensor-product operations in Nekbone can be optimized for two state-of-the-art GPU architectures: Pascal and Volta [6], [7], [8]. In addition, we evaluate the optimization compared to a measured roofline [9], [10] as to assess how close to peak performance we are.

## II. RELATED WORK

Previous work has been made on porting Nekbone to GPUs and regarding tensor-product operations related to SEM in general. In particular, Gong et al. [11] has made an initial GPU implementation of Nekbone and a pure OpenACC [12] version of Nek5000 [5]. In their work on Nekbone, it was noted that OpenACC could easily be used with the older Fortran code, but that compared to a CUDA [13] kernel for the main tensor-product operation, $Ax$, it did not perform as well. Therefore, they concluded that a trade-off between performance, ease of implementation, and portability could be achieved by computing most of the simpler vector operations with OpenACC and having a dedicated CUDA kernel for the tensor-product.

In parallel with this, work has been carried out by the CEED project on a refactored version of Nek5000 called NekRS [14]. The goal with NekRS is to create a modern Nek5000 and to utilize C++ instead of Fortran for the main bulk of the code, but to apply the OCCA library [15] to create kernels for various computing units [14], [16]. In part of this work, they optimized the tensor-product operation used in Nekbone with OCCA and assessed the performance gains. In their measurements, they achieved performance close to a measured roofline but did not use it in Nekbone itself [10].

In our work, the optimizations put forth by Świrydowicz et al. [10] are used in the original Nekbone application. We continue with the mixed OpenACC and CUDA approach initiated by Gong et al. [11] to interface with the original Fortran code.

## III. BACKGROUND TO NEKBONE

In this section, we introduce Nekbone and the tensor-product operation we will focus on in this work. Nekbone isolates the core computational components for Nek5000 by discretizing the Poisson equation with SEM on a cubic domain. It then uses the Conjugate Gradient (CG) method to solve the resulting linear system $Ax = f$. This approach is similar to Nek5000 and maintains the overall structure of the code. Most importantly, the evaluation of the Poisson operator in $Ax$, which is the most time-consuming part of Nek5000, is also part of Nekbone. Therefore, by optimizing $Ax$ in Nekbone we can show future performance improvements in Nek5000. We compute the $Ax$ subroutine in two steps. First, we express the local Poisson operator on each element as a tensor-product operation. Then, in the second step, we communicate the local computation results to the neighboring elements. We focus on optimizing the local tensor-products performed on each element. These tensor-products calculate the Poisson operator on the element as a series of matrix multiplications and depend on the differential operator's part, but also on the nodal values of the element and the geometry of the domain. The original approach is illustrated in Listing 1. In the pseudocode, we compute the local evaluation, $w$, of the Poisson operator for each element as a function of the nodal values `u`, the differential matrix `dxtm1` and the geometric factors `gxyz` similarly to how it is expressed in Nekbone. Note that the main operations are in essence small matrix multiplications, and thus we cannot use standard libraries such as cuBLAS [17] since they would not yield high performance.

Listing 1
PSEUDOCODE FOR THE TENSOR-PRODUCT USED TO EVALUATE THE POISSON OPERATOR IN NEKBONE.

```
do e = 1, nelt
    do i, j, k = 1, n   !Loop over i, j then k
        wr = 0
        ws = 0
        wt = 0
        do l = 1, n
            wr = wr + dxm1(i, l)*u(l, j, k, e)
            ws = ws + dxm1(j, l)*u(i, l, k, e)
            wt = wt + dxm1(k, l)*u(i, j, l, e)
        enddo
        ur(i, j, k, e) = gxyz(i, j, k, 1, e)*wr
                       + gxyz(i, j, k, 2, e)*ws
                       + gxyz(i, j, k, 3, e)*wt
        us(i, j, k, e) = gxyz(i, j, k, 2, e)*wr
                       + gxyz(i, j, k, 4, e)*ws
                       + gxyz(i, j, k, 5, e)*wt
        ut(i, j, k, e) = gxyz(i, j, k, 3, e)*wr
                       + gxyz(i, j, k, 5, e)*ws
                       + gxyz(i, j, k, 6, e)*wt
    enddo
enddo
do e = 1, nelt
    do i, j, k = 1, n   !Loop over i, j then k
        w(i, j, k, e) = 0.0
        do l = 1, n
            w(i, j, k, e) = w(i, j, k, e)
                          + dxtm1(i, l)*ur(l, j, k, e)
                          + dxtm1(j, l)*us(i, l, k, e)
                          + dxtm1(k, l)*ut(i, j, l, e)
        enddo
    enddo
enddo
```

### A. Cost Analysis

The cost of the tensor operations, together with other vector operations required for the CG method is

$$C(D, n) = D(12n + 34) \qquad (1)$$

per iteration, if we weight all floating point operations equally. We denote the degrees of freedom as $D$ and $n$ is the number of Gauss-Lobatto-Legendre (GLL) points in each dimension. The number of GLL points is related to the degree of our polynomial approximation as $n = $ degree $+ 1$. As for the bandwidth used, we execute $24D$ reads and $6D$ writes per CG iteration. We execute all our computations with double precision, so the computational intensity becomes

$$I(n) = \frac{D(12n + 34)}{8 \cdot 30D} = \frac{12n + 34}{240}. \qquad (2)$$

From equation (2) we can observe a higher computational intensity and also a higher performance for high polynomial degrees. For polynomial degrees used in real applications though, 7-9 [18], we expect the computation to be in the memory bound domain.

## IV. NEKBONE ON GPUS

In this section, we describe different implementations of Nekbone on GPUs as well as introduce the optimizations we have contributed with. In the current GPU implementation of Nekbone, the PGI compiler is used to compile the code on GPUs with a mixed OpenACC and CUDA approach [11]. OpenACC is a directive-based approach to GPU programming that can be used in the original Fortran code. It is used to run the majority of the vector additions as well as the communication on the GPU [12]. As for the Poisson operator, which we are optimizing, it has been approached in a few different ways.

### A. Original Implementation

In the original Nekbone implementation for GPUs by Gong et al. [11], the *Ax* subroutine (the Poisson operation) is implemented similarly to the pseudocode in Listing 1. Their original version suffers from a poor temporal locality, and only global memory is used. This initial approach has both a CUDA Fortran [19] and OpenACC implementation. In this subroutine, they allocate one thread block per element and then utilize as many threads as possible to compute the element with some stride (depending on the number of threads and size of the element). Despite the greedy use of available threads for a given element, they are not organized for locality. CUDA Fortran is a CUDA version developed by PGI and both CUDA C and CUDA Fortran can be called in the original Fortran code [19].

### B. Leveraging shared memory

Previous work focused on optimizing the kernel to exploit GPU shared memory [20]. Previous work achieved this by loading all the nodal values and the differential matrix into GPU shared memory and then executing the kernel as in the original approach. Using GPU shared memory enabled higher bandwidth and limit the global memory bandwidth used. However, since they utilized a 3D thread structure, where one thread is allocated for each nodal point in the element, they needed to load the entire element into shared memory. A problem with this is the limitation in GPU shared memory capacity and it prevents storage of the whole element, given a larger element size. For a P100 GPU this approach does not work for elements with more than 10 GLL points.

### C. 2D Thread Structure

Our main optimization is to utilize a new, 2D thread structure when creating our thread blocks. This thread structure avoids the problems in the shared memory version by only having a "layer" of points in shared memory at a time. The use of a different thread structure was studied in [10] and it enables other optimizations such as loop unrolling. What this 2D thread structure means is that instead of allocating as many threads as possible per element, each thread block instead uses a layer of $n \times n$ threads as shown in Fig. 1. By reorganizing the threads in this way, the outer two element-level loops in Listing 1, as well as the one iterating over $k$, can be merged. This means that each thread block goes through each $k$ layer in lock-step, enabling the threads to save `u, w` into registers as well as preloading the geometric factors, in addition to having `dxtm1` in GPU shared memory. When neighboring nodal values are needed for the second part of the differential, the threads in the block are synchronized (`__syncThreads()`). Our implementation trades inexpensive thread synchronization over the thread block (twice per layer) for increased temporal locality. We made this implementation in both CUDA C and CUDA Fortran. When using CUDA C, loop unrolling (`#pragma unroll`) as well as marking memory as read-only can be used for more compiler side optimizations. For our optimized CUDA Fortran version, the loop was manually unrolled once instead.

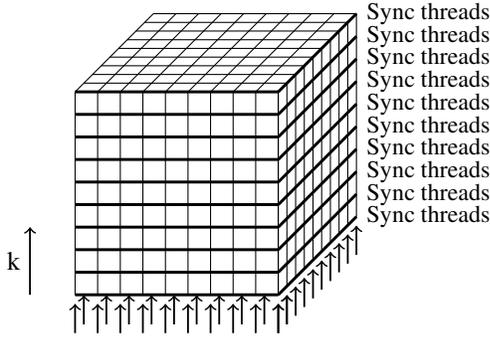

Fig. 1. Image illustrating the 2D thread structure for a thread block as a layer of arrows propagating in sync through an element and its k layers.

## V. EXPERIMENTAL METHODOLOGY

We perform experiments on Piz Daint at CSCS in Switzerland and Kebnekaise at HPC2N in Sweden. Piz Daint is equipped with Cray XC50 compute nodes and each node has a 12 core Intel Xeon E5-2690 v3 @ 2.6GHz and one Nvidia Tesla P100 GPU. Kebnekaise has a Intel Xeon Gold 6132 with 28 cores @ 2.6GHz and two Nvidia V100 GPUs per node. We use the PGI compiler 19.7 and CUDA version 10.1 on Piz Daint while on Kebnekaise we use version 18.7 and 9.2. Our compiler flags are `-acc` and `-Mcuda` to compile the OpenACC and CUDA code.

In all measurements, we run Nekbone with 100 CG iterations, polynomial degree 9 and for 10 GLL points in each dimension.

We conduct single GPU performance measurements at both Piz Daint and Kebnekaise. We measure the performance of the original CUDA Fortran version of the Poisson operator, the OpenACC version, the shared memory version by Gong et al., as well as our optimized CUDA Fortran and CUDA C implementations. Our measurements are performed with 64 - 4096 elements per GPU on Piz Daint. On Kebnekaise we make measurements with 448 - 3584 elements as to match a CPU node running with 16 - 128 elements per core (28 cores), which is the strong scaling interval for the current CPU code [18].

To evaluate how close we are to the peak performance of an Nvidia P100 and V100 GPU, we also measure the global memory bandwidth. We measure the bandwidth by running the same experiments as before, but instead of executing the computations, a `cudaMemcpy()` on the GPU is executed for each load and store in each CG iteration, excluding the communication between elements and nodes. This results in exactly double the amount of data movement necessary, and with this, a roofline for the performance can be calculated. We then compare the roofline to our optimized Nekbone implementation, but without the communication activated.

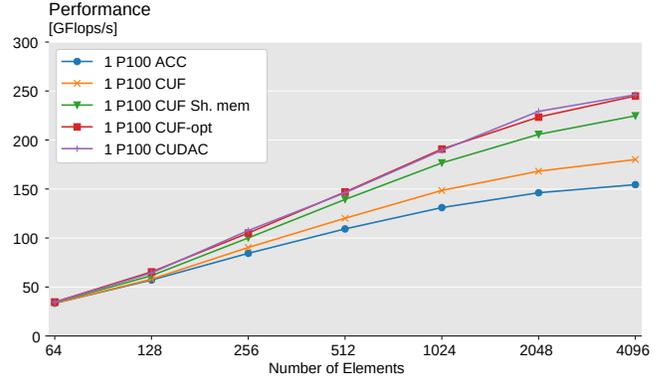

Fig. 2. Performance for different versions of Nekbone on Piz Daint with one Nvidia P100 GPU.

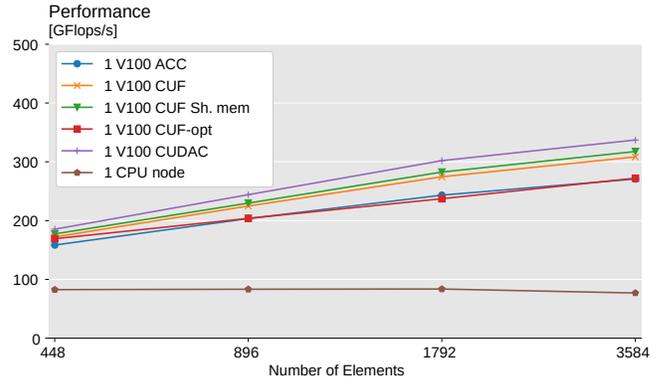

Fig. 3. Performance for different versions of Nekbone on Kebnekaise. The CPU version is run on one node with 28 cores and MPI for parallelization. The GPU measurements use one Nvidia V100 GPU.

## VI. RESULTS

In this section, we present and discuss the results of our performance measurements. We discuss the optimization and compare it to other versions, and then we make a comparison to a measured roofline. We have made our repository available online.[1] In all measurements, the results differed by less than 5% at most and error bars are therefore omitted.

### A. Performance of Optimization

The number of elements as well as bandwidth utilization impact our performance. By analyzing Fig. 2 we can observe the performance of all the major versions of Nekbone on a Nvidia P100 GPU while in Fig. 3 the performance is shown for a Nvidia V100. The restructuring of the kernel into 2D thread blocks gives improved performance. In particular, the difference between using global, shared, or register memory is noticeable. However, for the measurements on Nvidia V100 GPU, we do not observe any performance gain for the optimized CUDA Fortran kernel, but rather a slowdown. This can be attributed to the version of the PGI compiler used. Earlier measurements from Piz Daint, with an earlier compiler version, also show a large performance impact when using CUDA Fortran compared to CUDA C [20].

For later versions of the PGI compiler, however, the performance difference between CUDA Fortran and CUDA C

---
[1] https://github.com/MartinKarp/Nekbone/tree/cuda-openacc

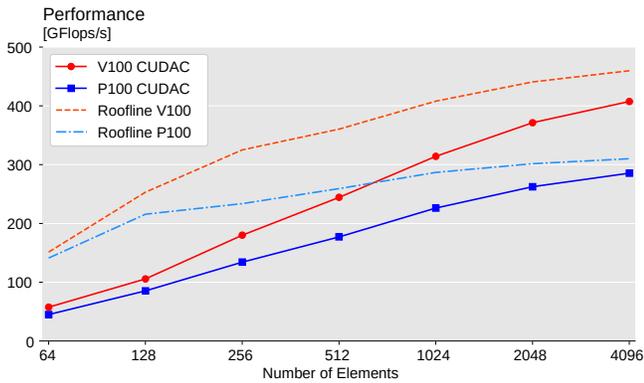

Fig. 4. Measured roofline for the P100 GPU on Piz Daint as well as the same measurements performed on a V100 GPU on Kebnekaise.

appears to be marginal. The performance difference between our optimized CUDA C and CUDA Fortran kernels is less than 1% on average on Piz Daint.

We include CPU performance for the Kebnekaise machine to illustrate the problem-size-dependence of the GPU version. This is a limiting factor for the scaling potential of GPUs. However, the debate between GPU vs. CPU is out of this paper's scope.

When comparing different GPU implementations, our implementation performs 10% better than the previous work's shared memory version and 36% better than the original approach on the Nvidia P100 GPU. For the Nvidia V100 GPU, we obtain a performance increase of on average 10% compared to the original version and 6% compared to the shared memory version. The GPU shared memory is combined with L1 cache on the V100 and the benefits of using it explicitly are therefore limited [7]. Our implementation is not bound by GPU shared memory and can, by only changing a few constants, be ported to other polynomial degrees than 9. This is an improvement compared to the shared memory approach, which does not work for larger elements. The question is then, how close to peak performance we are and how much higher performance can be achieved.

*B. Roofline*

To evaluate how close we are to peak performance, we construct a measured roofline [9], [10] in Fig. 4. These plots imply that we are in the memory-bound domain (Nekbone has been found to be memory bound for several computing architectures [21]). Based on the measurements of the bandwidth and our computational intensity (2), we note that we are close to the peak performance of the GPU for input sizes larger than 1024. For 1024, 2048, 4096 elements we achieve 78%, 87%, 92% of the roofline for the P100 and 77%, 84%,88% for the V100. We exclude smaller input sizes since the problem size then is too small and sensitive to kernel overhead.

The reason we use a measured roofline is that the bandwidth is dependent on the problem size. If we would be able to utilize the theoretical peak bandwidth of a P100 or a V100 GPU (720 GB/s and 900 GB/s respectively [6], [7]) we would according to (2) be able to achieve up to 462 GFlops/s for the P100 and 577 GFlops/s for a V100. The actual bandwidth we can utilize is lower and so is the performance we can achieve.

We observe a small performance difference when comparing the measured rooflines between the two GPU architectures. Our optimization of the tensor-product operation performs almost equally well compared to the respective rooflines for both GPUs, with a 1-4% better performance on the P100. It would appear that the performance of the original implementations for the V100 as shown in Fig. 3 is higher than the initial implementation on the P100 (Fig. 2), as a percentage of peak performance.

The reason we are not achieving performance on the measured roofline can be attributed to three main factors. Firstly, there is some kernel overhead, and this is especially clear for smaller inputs, but it decreases as the problem size increases. Secondly, most simpler operations are still made with OpenACC, such as the masking of boundary conditions. Masking values can affect performance since they are not well aligned in memory. Thirdly, since we are using `cudaMemcpy()` and not mimicking the actual read-write operations of Nekbone, our roofline is an optimistic bound for the application. With our mix of reads and writes in Nekbone we might not be able to utilize that amount of bandwidth.

## VII. CONCLUSION AND FUTURE WORK

In this paper, we have achieved a highly performant GPU version of Nekbone for single GPU for the Pascal and Volta GPU architectures that is not bound by shared memory. We do not expect to be able to optimize it a lot further since the performance was close to the measured roofline, according to our roofline analysis. Overall, using OpenACC for simpler operations did not appear to impact the code's performance by more than a few percentage points.

As for the performance, we notice that the performance on GPUs is sensitive to the change in input size. Because the performance for inputs smaller than 512 elements was so low, the scaling and in particular the strong scaling for smaller problem sizes will always be an issue compared to CPUs. Regardless of how communication between nodes affects the performance of Nekbone, we can conclude that having less than 500 000 degrees of freedom per GPU will not be beneficial for performance. In this case, increasing the number of elements on the GPUs will increase performance almost as much as using more GPUs.

In the future, we will investigate the communication kernel and the preconditioned CG method. In real applications, the use of a preconditioner can improve the convergence and reduce the run-time by several orders of magnitude, so running without one is seldom done. We expect this to be an issue for GPUs because several preconditioners have a lower computational intensity and more communication is involved per iteration [22].

ACKNOWLEDGMENT

This work was supported by a European Commission Horizon 2020 project grant entitled "EXCELLERAT: The European Centre of Excellence for Engineering Applications" (grant reference 823691). The computations were performed on resources provided by the PRACE Research Infrastructure resource Piz Daint hosted by CSCS, Switzerland and the Swedish National Infrastructure for Computing (SNIC) at High Performance Computing Center North (HPC2N).